\RequirePackage[2020-02-02]{latexrelease}
\documentclass[aps,cyrwin,prb,twocolumn,floats,eqsecnum,graphicx,epsfig,euscript,amsmath,amssymb]{revtex4}
\usepackage{mathrsfs}
\usepackage{graphicx}
\usepackage{epsfig}

\DeclareMathOperator{\Rot}{rot}
\DeclareMathAlphabet\EuScript{U}{eus}{m}{n} \SetMathAlphabet\EuScript{bold}{U}{eus}{b}{n}

\renewcommand{\min}{\mathop{\rm min}\nolimits}
\renewcommand{\max}{\mathop{\rm max}\nolimits}
\def\lapprox{\,\raise0.4ex\hbox{$<$}\kern-0.8em\lower0.7ex\hbox{$\sim$}\,}
\def\gapprox{\,\raise0.4ex\hbox{$>$}\kern-0.8em\lower0.7ex\hbox{$\sim$}\,}
\def\be{\begin{equation}}
\def\ee{\end{equation}}
\def\ba{\begin{eqnarray}}
\def\ea{\end{eqnarray}}

\def\bc{\begin{center}}
\def\ec{\end{center}}

\begin{document}

\title{Massive spin-flip excitations in a \mbox{\boldmath{$\nu\!=\!2$}} quantum Hall ferromagnet}

\author{S. Dickmann and P.$\,$S. Berezhnoy }
\affiliation{Institute of Solid State Physics, RAS, Chernogolovka, 142432, Moscow
District, Russia}

\date{\today}

\begin{abstract}
{Excitation with a massive spin reversal of the individual skyrmion/antiskyrmion type is theoretically studied in a quantum Hall ferromagnet, where the zero and first Landau levels are completely occupied only by electrons with spins aligned strictly in the direction
determined by the magnetic field. The Wigner-Seitz parameter is not necessarily considered to be small. The microscopic model in use is based on a reduced basic set of quantum states [the so-called ``single-mode (single-exciton) approximation''], which allows proper account to be taken for mixing of Landau levels, and substantiating the equations of the classical $O(3)$ nonlinear $\sigma$ model. The calculated `spin stiffness' determines the exchange gap for creating a pair of skyrmion and antiskyrmion. This gap is significantly smaller than the doubled cyclotron energy and the characteristic electron-electron correlation energy. Besides, the skyrmion--antiskyrmion creation gap is much smaller than the energy of creation of a separated electron--exchange-hole pair calculated in the limit case of a spin magnetoexciton corresponding to an infinitely large 2D momentum. At a certain magnetic field (related to the 2D
electron density in the case of fixed filling factor $\nu$), the gap vanishes,
which presumably points to a Stoner transition of the quantum Hall ferromagnet to a paramagnetic phase.}
\end{abstract}

\vspace{1mm}

\maketitle

\bibliographystyle{prsty}

\vspace{-4mm}

\section{Introduction}
Interest in massive spin excitations in quantum-Hall (QH) ferromagnets, where
change of total spin, $\delta S$, is large ($|\delta S|\!>\!1$) was triggered
by the pioneering theoretical work of Sondhi {\it et al};\cite{so93}
and intensified after the experimental discovery of a massive spin flip near
the ground state of a QH ferromagnet.\cite{ba95} This interest
is also due to the fact that, with increase in the number of inverted spins, the
excitation energy, being of exchange origin, decreases.\cite{pa96} So,
according to the theory, at the $\nu\!=\!1$ filling factor and in the ideally strict two-dimensional (2D) case, the energy gap of creation of a skyrmion-antiskyrmion
pair (where $|\delta S|\!\gg\!1$) is approximately by half less than the gap for an
electron--exchange-hole pair (where $|\delta S|\!=\!1$),\cite{so93} provided the Zeeman
energy is neglected. Thus, the study
of excitations with a massive spin flip turns out to be closely related to the
problem of the size of the activation gap in QH transport.

The present paper is devoted to the QH ferromagnet, where equally
spin-polarized electrons occupy both zeroth and
first lower Landau levels, and, due to the peculiarities of the real $\nu\!=2$
ferromagnet,\cite{ma14} the Wigner-Seitz parameter of the system
is not necessarily considered to be small. In this introductory section, we
present some reasons for the model in question.

It is well known that the interparticle correlations in a multi electronic
ensemble  are responsible for the most interesting properties of quantum Hall
systems (QHSs). The interaction is usually characterized by the Wigner-Seitz
 parameter $r_s$ that in QHSs with fixed filling factor $\nu\!\sim\!1$, is
 in fact the ratio of the characteristic Coulomb energy,
 ${\mathscr E}_{\rm C}\!=\!(e^2/\kappa)\sqrt{n_{\rm s}}$, to the cyclotron one,
 $\hbar\omega_c$. (Here $n_{\rm s}$ is the 2D electron density related to the magnetic
 field by equation $n_s\approx 2.4\times 10^{10}\cdot\nu\cdot B[T]/$cm${}^2$; $\kappa$ is the
 dielectric constant).
 Experimental and theoretical studies in the field of QHS physics are distinguished
 by a completely different relation to the $r_s$ magnitude. On the one hand,
 experiments investigating clean QH systems with a large $r_s$ value, such as, for
 example, ZnO/MgZnO structures,\cite{ma14,va17} demonstrate quite spectacular
 results and allow, in addition to `classical' quantum Hall phenomena (e.g., the
 features of the $\nu$-fractional transport), to discover even new effects. The
 latter include, for instance, Stoner magnetic transitions.\cite{ma14,va17,di20} On
 the other hand,
 theory, due to impossibility to use some perturbative technique based on the
 $r_s$ smallness (cf. works in Ref. \onlinecite{by81}), hardly `copes' with the study of QH
 systems with large $r_s$. In this situation, there are two different theoretical
 approaches. The first is presented by numerical calculations (numerical experiments), where a
 fairly limited number of interacting electrons is considered. It also involves controversial
 assumptions about a certain decrease in the effective Coulomb constant.\cite{lu16}
 The other approach is represented by the well-known studies that use conceptually new semi-fenomenological
 models to, at least indirectly, account for strong correlation in the electronic continuum
 (see, for example, the milestone works
 \onlinecite{la84}). The Landau level mixing is, however, ignored even in these
 fairly successful studies, despite the fact that the experimental value of $r_s$  is not very
 small ($r_s\!\simeq\!0.3-0.7$ in GaAs/AlGaAs structures).

At large $r_s$ (e.g., when $1\!\lesssim\! r_s\!\lesssim\,$10) we can assume that
the electron distribution is effectively smeared across a dozen Landau levels. There are,
however, reasons to believe that such a notion is wrong. Indeed, a smearing
distribution over Landau levels can hardly be compatible with well observed (even
at large $r_s$) sharply non-monotonic dependencies of the transport and optical
properties on the value of the filling factor $\nu$. On the contrary, there are
many signs indicating that the Fermi-liquid paradigm is valid also for the system studied.
That is, the strong interaction retains the same classification of energy levels
as in the ideal Fermi gas, yet, leads to a renormalization, presenting
interacting particles/electrons as quasiparticles with definite momenta.\cite{llv9} The
distribution of quasiparticles over momenta at $T\!\to\!0$ is the step-function,
$\theta(p_F\!-p)$, where the 2D Fermi momentum, $p_F$, is defined in terms of
the total density of particles (or, what is the same, quasiparticles) by the usual expression:
$p_F/\hbar\!=\!(2\pi n_{\rm s})^{1/2}$. As for the distribution of true particles over momenta, it
is certainly not represented by the strict step-function above, yet, at any number
$r_s$ has also a discontinuity at $p\!=\!p_F$ (the so-called Migdal
jump$\,$\cite{llv9,mi57}) and the interaction results only in some tails in the
distribution at $p\!>\!p_F$ (electrons) and at $p\!<\!p_F$ (`holes').
Thus, in a QHS there should be no very smooth distribution of electrons over the
Landau levels in the vicinity of the Fermi energy. The Fermi-liquid picture for describing the 2D electronic system is supported, for instance, by the results of the recent study that has been carried out in zero and weak magnetic fields at temperatures
$T\!\simeq\!25\,$mK.\cite{ku22} Typical Fermi-liquid features (in particular,
observation of the Migdal jump), were found under the conditions corresponding to values $r_s\!=\!{\mathscr E}_{\rm C}/E_F\!\simeq\!4.5$ (the Fermi energy is
$E_F\!=\!p_F^2\!/2m_e^*$). 

Our goal is to find the excitation energy from the
ground state, which we will model using a step function, regardless of the
value of parameter $r_s$. So, the QH ferromagnet at $\nu\!=$2 is
considered as a system where the states with spins `pointing up' at the
zero and first Landau levels are fully occupied, and all other states are
completely empty. Of course, this picture can be interpreted as a
renormalization, i.e. a transition to the concept of Fermi-liquid
quasiparticles. However, if we assume that the cyclotron gap is larger than
the lengths of the energy `tails' in the distribution of real electrons, we may
not actually make a difference between the Fermi-liquid quasiparticles and the
electrons.

Besides, our presumably extensive (spatially smooth) spin excitation makes it
possible to use a perturbative approach based on the smallness of the spatial
derivatives of the spin-rotation matrix components. (The part of the Schr\"odinger
operator responsible for the
interaction is invariant to spin-rotation, so such derivatives appear
only due to the action of the single-particle Schr\"odinger operator on the
spin-rotation matrix.) The smooth rotation enables us to consider the multi
electronic Schr\"odinger
equation separately on two different spatial scales: on the scale of the
spatial change of the local spin, and that of the
change of the electron wave function. The former is determined by the spatial size (core) of
the skyrmion, which is determined by the small Zeeman$/\!$exchange energy ratio (see, e.g., Ref. \onlinecite{by98}); the latter is the magnetic length,
$l_B$. When considering
a domain with a dimension much less than the skyrmion size but larger
than $l_B$, one can study a `local' QHS represented by a domain perturbed by a
weak gauge field, homogeneous within the chosen domain, which
is added to the vector potential.\cite{io99,di02} In particular, some `fake'
magnetic field arises, slightly renormalizing the cyclotron energy and
magnetic length. Calculating the correction to the
energy of this ferromagnetic domain, we use a model of the reduced basic-set
describing the QHS states. This is so-called single-mode$\,$\cite{ma85} or, in
other words, a single-exciton approximation (see, e.g., Ref. \onlinecite{di20}).
Owing to the homogeneity of the
perturbating field, the relevant single-exciton basic set contains only states
that do not violate the translation symmetry of the system, i.e. represent
only `vertical' mixing of the Landau levels. The basis set consists of
certain combinations of electron promotions from one Landau level to
another occurring with or without a spin flip.

It is clear that the energy determined by rotation of the spins in the space
is vanishing in the case of a zero Zeeman gap and zero spin stiffness
(i.e. at zero exchange energy associated with a spin flip). Indeed,
then any spin rotation actually becomes single-electronic, and at a zero
Zeeman gap it is certainly gapless. Zero spin stiffness occurs when, at
fixed filling factor $\nu$, the $r_s$ parameter goes to zero (formally, this can be
achieved if the dielectric constant of the lattice goes to infinity,
$\kappa\!\to\!\infty$).

Second, interestingly, in the opposite extreme case, when the stiffness is
infinitely large (for fixed $\nu$, it means that $r_s\!\to\!\infty$),
the exchange corrections to the energy found for a small domain can be also predicted
to be vanishing. Indeed, these represent the second-order corrections calculated perturbatively
in terms of a weak magnitude of the additional field proportional to the small gradients of the
spin-rotation matrix components. They are of two kinds: occurring
due to mixing of the ground state with zero-momentum magnetoplasma modes
without any spin change; or appearing as a result of mixing
with zero-momentum  cyclotron--spin-flip modes. The latter contribute only to the
second-order correction, determined by the terms with denominators
containing large exchange energies of the order of ${\mathscr E}_{\rm C}\!\gg \!\omega_c$.
These terms are vanishing if $r_s\!\to\!\infty$, and the main contribution to the excitation
energy is due to mixing with soft spinless magnetoplasma states.
Thus, the transport gap actually becomes of the order of cyclotron energy
$\omega_c$, being much smaller than ${\mathscr E}_{\rm C}$. At a fixed total number
of electrons, the gap corresponds to the excitation of a pair consisting of an individual skyrmion and an anti-skyrmion.

\section{Energy of massive spin excitation}

\subsection{Smooth rotation in the spin space}

We use the approach similar to that described in previous studies$\,$ \cite{io99,di02}.
The rotation of the electron spins in the 3D space is determined by the rotation matrix
${\hat U}(\mbox{{\boldmath $R$}})\,$ \cite{ll91} parametrized by three Eulerian
angles $\alpha(\mbox{{\boldmath $R$}})$, $\beta(\mbox{{\boldmath $R$}})$, and
$\gamma(\mbox{{\boldmath $R$}})$ (see also Appendix B below) smoothly depending on the
2D spatial coordinate
$\mbox{{\boldmath $R$}}\!=\!(X,Y)$. Only two of the angles, $\alpha$ and $\beta$, are
sufficient to fully determine the local direction of the spin described by the unit
vector of $\vec{\, n}(\mbox{{\boldmath $R$}})$. Operator ${\hat U}$ represents a $2\times 2$ matrix
transforming the spinor $\vec{\, \psi}(\mbox{{\boldmath $R$}})$ given in the spatially
fixed system $\{{\hat x},{\hat y},{\hat z}\}$ to a `local' spinor
$\vec{\,\chi}(\mbox{{\boldmath $R$}})$, given in the `local' coordinate system
$\{{\hat x}',{\hat y}',{\hat z}'\}$, and thus accompanying the spatial
spin-rotation [so that, for instance, we always have $\langle\vec{\,\chi}\rangle\propto
{1\choose 0}$,
where brackets $\langle...\rangle$ mean averaging over a small domain in the
vicinity of the
$\mbox{{\boldmath $R$}}$ coordinate]. That is,\vspace{-1mm}
\begin{equation}\label{rotation}
\vec{\, \psi}(\mbox{{\boldmath $R$}})\!=\!{\hat U}
(\mbox{{\boldmath $R$}})\vec{\, \chi}(\mbox{{\boldmath $R$}}).\vspace{-1.mm}
\end{equation}
It is convenient to choose ${\hat z}$ as the Zeeman axis, i.e. to
consider the magnetic field parallel to ${\hat z}$ in the fixed coordinate system,
but oppositely directed (assuming for certainty the Land\'e factor to be
positive, that is $\vec{B}\cdot{\hat z}=-B$). In the case of a
tilted magnetic field, the Zeeman axis ${\hat z}$ is inclined by a constant angle
$\theta$ with respect to the quantization axis ${\hat Z}$, where ${\hat Z}$ is
perpendicular to the $\{{\hat X},{\hat Y}\}$ plane. (The coordinate systems
$\{{\hat x},{\hat y},{\hat z}\}$ and $\{{\hat x}',{\hat y}',{\hat z}'\}$, used
for description of the spin-orientation, should not be confused with the coordinate
system $\{{\hat X},{\hat Y}\!,{\hat Z}\}$ in the real space, in which the 2D
radius-vector $\mbox{{\boldmath $R$}}$ indicates the coordinates on the plane
$\{{\hat X},{\hat Y}\}$.)

We will be looking for the $U(\mbox{{\boldmath $R$}})$ matrix that
satisfies certain conditions, namely, the unit vector
\vspace{-2.mm}
\begin{equation}\label{unit_n}
{\vec{\, n}}(\mbox{{\boldmath $R$}})\propto\,\langle{\!\vec{{}\,\psi}}\,{}^\dag\!
(\mbox{{\boldmath $R$}})
\widehat{\mbox{{\boldmath $S$}}}{{\vec{{}\,\psi}}(\mbox{{\boldmath $R$}})}
\rangle\vspace{-2.mm}
\end{equation}
($\widehat{\mbox{{\boldmath $S$}}}$ stands for the operator of the total spin),
specified in the $\{{\hat x},{\hat y},{\hat z}\}$ space and indicating the local
spin orientation, should not have any singularities at any $\mbox{{\boldmath $R$}}$, and
should have a fixed $z$-orientation in the core (i.e. at $\mbox{{\boldmath $R$}}\!=\!0$) and on
the periphery
(at $|\mbox{{\boldmath $R$}}|\!\to\infty$). This means that
$\beta(0)\!=\!\pi$ and $\left.\beta(\mbox{{\boldmath $R$}})\right|_{R\!\to\infty}\!\!=\!0$
regardless of the azimuth given by value $\alpha(\mbox{{\boldmath $R$}})|_{R\!\to\infty}$.
Substituting $\langle\vec \chi\rangle\!=\!{1\choose 0}$ into Eq.
\ref{rotation}, and taking into account Eq. \ref{unit_n}, we obtain
expressions of the components of the unit vector in terms of the Euler angles
$\alpha(\mbox{{\boldmath $R$}})$ and $\beta(\mbox{{\boldmath $R$}})$,\vspace{-1.mm}
\begin{equation}\label{vec_n} n_x\!=\sin{\!\beta}\cos{\alpha},\,\,
n_y=\sin{\!\beta}\sin{\alpha},\,\, n_z\!=\cos{\!\beta}.\vspace{-1.mm}
\end{equation}
Rotation of each spin around the Zeeman axis at the same angle leaves unchanged
energy and other quantities that have a physical meaning. That is, there is
invariance with respect to transition\vspace{-2.mm}
\begin{equation}\label{phi}
\alpha(\mbox{{\boldmath $R$}}) \to \alpha(\mbox{{\boldmath $R$}})+ \phi,
\vspace{-1.5mm}
\end{equation}
where $\phi$ is a constant independent of $\mbox{{\boldmath $R$}}$. In particular,
in the fixed coordinate system $\{{\hat x},{\hat y},{\hat z}\}$ the value $\phi\!=\!0$
corresponds to a `radial' rotation of vector $\vec{\, n}(\mbox{{\boldmath $R$}})$
(a Neel-type skyrmion), whereas, e.g.,
$\phi\!=\!\pi/2$ means a `tangential' rotation (a Bloch-type skyrmion).

The ${\hat y}$ axis of the coordinate system $\{{\hat x},{\hat y},
{\hat z}\}$ and the ${\hat Y}$ axis of $\{{\hat X},{\hat Y}\!,{\hat
Z}\}$ can be always chosen to coincide with the line of intersection of the planes
$\{{\hat x},{\hat y}\}$ and $\{{\hat X},{\hat Y}\}$. Hence, both coordinate
systems are combined by turning of the $\{{\hat x},{\hat y},
{\hat z}\}$ system at angle $\theta\!=\!\widehat{({\hat z},{\hat Z})}$ around
${\hat Y}$. As a result, the unit vector ${\vec{\,n}}=(n_x,n_y,n_z)$ in the system
$\{{\hat X},{\hat Y}\!,{\hat Z}\}$ is presented by components
\vspace{-1.mm}
\begin{equation}\label{vec_i}
{}\!{}\!{}\!\begin{array}{l}
{}\!{}\!n_X\!\!=\!n_x\!\cos{\theta}\!+\!n_z\sin{\theta},\;\; n_Z\!\!=\!n_z\!\cos{\theta}\!-
\!\!n_x\sin{\theta},\vspace{0.0mm}\\{}\!{}\!n_Y=n_y.
\end{array}\vspace{-1.mm}
\end{equation}

Macroscopically, the magnet energy of the 3D unit vector \ref{vec_n} is described
in the framework of an $O(3)$ nonlinear $\sigma$ (NL$\sigma$) model (see Appendix A),
whose equations can be substantiated microscopically in the case of a quantum Hall
ferromagnet.

\vspace{-2.mm}
\subsection{Microscopic approach; Hamiltonian}
\vspace{-2.mm}
Before describing our system microscopically, we draw attention to the
hierarchy of distance scales. The scale of the wave function is determined by the
magnetic length, which is assumed to be much smaller than the
spatial scale of the spin change (let the latter be designated as $\Lambda$, that is,
$l_B\!\ll\!\Lambda$ where
$\Lambda$ is the size of the ``skyrmion
core'', see the next section and Appendix A). We study a single excitation, therefore, $\Lambda$ is
considered to be much smaller than the mean distance between excitations in the
system in question. In addition, if the excitation is charged, the parameter
$\Lambda$ characterizes the spatial change in the charge density.

In the following, for every coordinate $\mbox{{\boldmath $R$}}$ we use the
substitution:\vspace{-2.mm}
\begin{equation}\label{r}
\mbox{{\boldmath $R$}}\to \mbox{{\boldmath $R$}}+
\mbox{{\boldmath $r$}},\vspace{-2.mm}
\end{equation}
where $\mbox{{\boldmath $r$}}$ belongs to a small
domain ${G}$${}\!\!{}_{\tiny{\mbox{\boldmath $R$}}}$ in the vicinity of point
$\mbox{{\boldmath $R$}}$. The domain area,
$\Delta^2\!\mbox{{\boldmath $R$}}\!=\!\Delta X\Delta Y$,  is considered to be much smaller than
$\Lambda^2$ (i.e. always $r\ll\Lambda$), however, let it be still considered much
larger than $l_B^2$. So, the integration over $\mbox{{\boldmath $R$}}$ can be
presented as summation over small domains $\Delta^2\!\mbox{{\boldmath $R$}}_i$ :\vspace{-1.mm}
\begin{equation}\label{dR}
\int...\,\,d^2\!\mbox{{\boldmath $R$}}\to \sum_i...\, \Delta^2\!\mbox{{\boldmath $R$}}_i\equiv
\sum_i\!
\!\!\!\!\!\!\!\!\!\int\limits_{\qquad\mbox{\scriptsize over\:}
G\!{}_{\tiny{\mbox{\boldmath $R$}}_i}}
\!\!\!\!\!\!\!\!\!\!\!...\,d^2\!\mbox{{\boldmath $r$}}.\vspace{-1.mm}
\end{equation}
We will present the entire area of our system as consisting of
$\textstyle{G}\!$${}_{\tiny{\mbox{\boldmath $R$}}_i}$ domains whose areas obey the condition:
$l_B^2\!\ll\,\mbox{area-of-}G$$\!{}_{\footnotesize{\mbox{\boldmath $R$}}_i}\ll$$\,\Lambda^2$;
so that
integration of a function $F(\mbox{{\boldmath $R$}})$ over the 2D space becomes
summation over the domains covering the total area of the system.\cite{foot2}

We start from the QHS Hamiltonian
\begin{equation}\label{Ham}
{\hat H}_{\rm tot}={\hat H}_{\rm Z}+
{\hat H}_{1}+{\hat H}_{\rm int},
\end{equation}
where\vspace{-1.mm}
\begin{equation}\label{Z_Ham}
{\hat H}_{\rm Z}=-\epsilon_{\rm Z}\!\int\!\! d^2\!\!\mbox{{\boldmath $R$}}
{\vec{{}\,\psi}}\,{}^\dag\!
(\mbox{{\boldmath $R$}})\,\hat{S}_z
\vec{{}\,\psi}(\mbox{{\boldmath $R$}})\vspace{-2.mm}
\end{equation}
stands for the Zeeman energy ($\epsilon_{\rm Z}\!=\!|g\mu_B\vec{B}|$ is the Zeeman
gap, $\vec{{}\,\psi}$ is the Schr\"odinger operator now);
${\hat H}_{1}$
represents the 2D `kinetic energy' term:
\begin{equation}\label{H1}
{\hat H}_1=\frac{1}{2m_e^*}\int\!d^2\!\!\mbox{{\boldmath $R$}}\,
  {\vec{\,\psi}}{}^\dag\!(\mbox{{\boldmath $R$}})
  \left(-i\mbox{{\boldmath $\nabla$}}+\mbox{{\boldmath $A$}}\right)^2\!\!
  {\vec{\,\psi}}(\mbox{{\boldmath $R$}})
\end{equation}
[$\mbox{{\boldmath $A$}}(\mbox{{\boldmath $R$}})$ is the vector potential, for instance:
$A_Y\!=\!XB_\perp\!\equiv\!XB\cos\theta,\,\,A_x\!\!=\!A_z\!\equiv 0$; besides,
we use units where $\hbar=e/c=1$, so that the cyclotron energy is $\omega_c=
B_\perp\!/m_e^*$]; and the electron interaction term is \vspace{1.mm}

\noindent$\large{\hat{ H}_{\mbox{\scriptsize int}}}$
\vspace{-.5mm}
\begin{equation}\label{Coul}
 {}\!\!\!{}\!{}\!{}{}\!{}\!{}=\!\!\displaystyle{\!\int\!\!\!\!\int\!\!\frac{d^2\!\!\mbox{{\boldmath $R$}}\,
  d^2\!\!\mbox{{\boldmath $R$}}'}{2}\!\vec{\, \psi}{}^\dag\!(\!\mbox{{\boldmath $R$}})\!
  \vec{\, \psi}{}^\dag\!(\!\mbox{{\boldmath $R$}}')V\!
  (|\!\mbox{{\boldmath $R$}}\!-\!\!\mbox{{\boldmath $R$}}'|)\!
  \vec{\, \psi}(\!\mbox{{\boldmath $R$}}'\!)\!\vec{\, \psi}(\!\mbox{{\boldmath $R$}})}
  \vspace{-1.5mm}
\end{equation}
[$V(R)$ is the Coulomb interaction vertex; as usual,
it is appropriately renormalized by taking into account
nonideal two-dimensionality of the electron system].

First, we focus on integration over the
$\textstyle{G}\!$${}_{\tiny{\mbox{\boldmath $R$}}}$ domain in the vicinity of
fixed point
$\mbox{\boldmath $R$}$. Substituting ${\hat S}_z\!=\!{\hat \sigma}_z\!/2$ and Eq.
\ref{rotation} into Eq. \ref{Z_Ham}, and remembering
that by definition we have $\vec{\, \chi}|0\rangle\propto{1\choose 0}\,|0\rangle$
(where $|0\rangle$ is
the ground state of the domain in the vicinity of $\mbox{\boldmath $R$}$), we
obtain the contribution of the
$\textstyle{G}\!$${}_{\tiny{\mbox{\boldmath $R$}}}$ domain to the Zeeman
energy: \vspace{-2.mm}
\begin{equation}\label{ZeemanG}
{}\!{}\!{}\!{}\!\displaystyle{\langle\hat{\cal H}_{\rm Z}(\mbox{\boldmath $R$})\rangle\!=\!-\!
\frac{\epsilon_{\rm Z}}{2}\cos{[\beta(\mbox{\boldmath $R$})]}\!\!\int\limits_{G\!{}_{\tiny{
\mbox{\boldmath $R$}}}}\!\langle\!\vec{\, \chi}{}\,^\dag\!\!\!{}_{\!\tiny{\mbox{\boldmath $R$}}}
 (\mbox{\boldmath $r$})\vec{\, \chi}\!{}_{\tiny{\mbox{\boldmath $R$}}}(
  \mbox{\boldmath $r$}\,)\rangle d^2\!\!\mbox{\boldmath $r$}}.\vspace{-3.mm}
\end{equation}
The `kinetic energy' term takes the form ${}\,$\cite{io99,di02}
\vspace{-2.mm}
\begin{equation}\label{H_1omega}
\!\begin{array}{l}
\displaystyle{\hat{\cal H}_1\!(\mbox{\boldmath $R$})\!=\!\frac{1}{2m_e^*}\!
\int\limits_{G\!{}_{{\tiny\mbox{\boldmath $R$}}}}\!\!d^2\mbox{\boldmath $r$}
 \,\vec{\, \chi}{}\,^\dag\!\!\!{}_{\!\tiny{\mbox{\boldmath $R$}}}
 (\mbox{\boldmath $r$})
  \!\left[-i\mbox{\boldmath $\nabla$}\!{}_{{\scriptsize\mbox{\boldmath $r$}}}
  \!+\vphantom{{\vec \Omega}{}^{(l)}\!(\mbox{\boldmath $R$})}
  {\mbox{\boldmath $A$}}_{\!\tiny{\mbox{\boldmath $R$}}}
  (\mbox{\boldmath $r$})\right.}\vspace{-3.mm}\\
\qquad\qquad\qquad\qquad\displaystyle{\left.\!+\,
{\textstyle \sum_{\, l}}\vec{\,\Omega}_{\!\tiny{\mbox{\boldmath $R$}}}\!\!\!{}^{(l)}
\!(\mbox{\boldmath $r$})\,
\hat{\sigma}_l\right]^2\!\!
  \vec{\, \chi}\!{}_{\tiny{\mbox{\boldmath $R$}}}(
  \mbox{\boldmath $r$})}.
\end{array}  \vspace{-1.mm}
\end{equation}
We make denotation corresponding to the replacement:
$\mbox{{\boldmath $A$}}\left(\mbox{{\boldmath $R$}}+\mbox{{\boldmath $r$}}\right)\to$
${\mbox{\boldmath $A$}}_{\!\tiny{\mbox{\boldmath $R$}}}
\left(\mbox{{\boldmath $r$}}\right)$, and the same for
$\vec{\, \chi}(\mbox{{\boldmath $R$}}+\mbox{{\boldmath $r$}})\,$ and
$\vec{\,\Omega}{}^{(l)}
(\!\mbox{{\boldmath $R$}}+\mbox{{\boldmath $r$}})$. Here $\;l=x,\,y,\,z\,$;
$\,\;{\hat \sigma}_{l}$ stands for Pauli matrices.
The 2D vectors
$\vec{\, \Omega}$${}_{\tiny{\mbox{\boldmath $R$}}}$${}\!\!\!{}^{(l)}$
${}\!\!\!(\mbox{{\boldmath $r$}})$ with components
$\!\vphantom{\Omega^{(l)}}\Omega^{(l)}$${}_{\!\!\!\!\!\tiny{\!\mbox{\boldmath $R$}},
\mbox{\scriptsize $x$}}$ and $\Omega^{(l)}{}_{\!\!\!\!\!\!\!\!\tiny{\mbox{\boldmath $R$}},y}\!
\vphantom{\Omega^{(l)}\!}\,$
are proportional
to the small spatial derivatives of the rotation-matrix components$\,$\cite{io99}
[see Eq. \ref{A_Omega} in Appendix B]. In fact,
only the values
$\vec{\, \Omega}$${}\!{}_{{}\tiny{\mbox{\boldmath $R$}}}$$\vphantom{
\vec{\, \Omega}^{(l)}}\!\!{}^{(l)}$${}\!
(0)$, and $\partial_{{}\mu}\!\!\!\left.\vec{\, \Omega}^{(l)}_{\!\tiny{\mbox{\boldmath $R$}}}
(\mbox{{\boldmath $r$}})\!\right|$${\vphantom{\left.\vec{\, \Omega}^{(l)}_{\!\tiny{\mbox{\boldmath $R$}}}
\!\!(\mbox{{\boldmath $r$}})\!\right|{}}}_{\mbox{\small{\boldmath $r$}}\to 0}\!{}$ ${}\!$
independent of $\mbox{{\boldmath $r$}}$ are essential within our approach (here $\partial_{{}\mu}
\!\!\equiv\!
\nabla_{{}\!\!\mu}$, where $\mu\!=x$ or $y$).

Finally, the form of interaction
term \ref{Coul} is simply invariant with respect to the rotational
transformation. By substituting Eq. \ref{rotation} into \ref{Coul}, for the
$\textstyle{G}$${}\!{}_{\tiny{\mbox{\boldmath $R$}}}$ domain:\vspace{1.mm} we find

\noindent$\hat{\cal H}_{\mbox{\scriptsize int}}(\mbox{\boldmath $R$})
\qquad\qquad\qquad\qquad\qquad\qquad$
\vspace{-2.mm}
\begin{equation}\label{CoulR}
{}\!\!\!\!{}\!{}=\!\!\!\!{}\!{}\!{}\!\displaystyle{\!\!{}\!\iint
\limits_{\quad\,\mbox{\scriptsize{\boldmath $r$}},
 \mbox{\scriptsize{\boldmath $r$}}'\in\,G\!{}_{\scriptsize{\mbox{\boldmath $R$}}}}\!\!
 \!\!\!{}\!{}\!{}\!{}\!{}\frac{d^2\!\mbox{{\boldmath $r$}}
  d^2\!\mbox{{\boldmath $r$}}'}{2}\!\vec{\, \chi}{}^\dag\!\!\!{}_{\!\tiny{\mbox{\boldmath $R$}}}
  (\mbox{{\boldmath $r$}}\!)\!
  \vec{\, \chi}{}^\dag\!\!\!{}_{\!\tiny{\mbox{\boldmath $R$}}}(\mbox{{\boldmath $r$}}')V\!
  (|\mbox{{\boldmath $r$}}\!-\!\mbox{{\boldmath $r$}}'|\!)\!
  \vec{\, \chi}\!{}_{\tiny{\mbox{\boldmath $R$}}}(\mbox{{\boldmath $r$}}'\!)\!
  \vec{\, \chi}\!{}_{\tiny{\mbox{\boldmath $R$}}}(\mbox{{\boldmath $r$}})}.
  \vspace{-1.mm}
\end{equation}

So, if not considering the Zeeman energy, then the
`$\vec{\,\Omega}^{(l)}\!\!{\hat \sigma}_l$'-terms in Eq. \ref{H_1omega} represent the only
thing that essentially distinguishes our
Hamiltonian describing the electrons of the
$\textstyle{G}\!$${}_{\tiny{\mbox{\boldmath $R$}}}$ domain from
the Hamiltonian that characterizes the system without any spin rotation. If we consider
that actually by definition we get
$\langle\!\vec{\,\chi}^\dag{\hat \sigma}_l\vec{\,\chi}\rangle\!=\!\delta_{l,z}$ and
$\langle\!\vec{\, \chi}^\dag{\hat \sigma}_l{\hat \sigma}_{l'}\!\vec{\, \chi}\rangle\!=
\!\delta_{l,z}\delta_{l'\!\!,z}$ (here and further $\delta_{...}$ is the Kronecker
delta), then
these terms lead to appearance of additional gauge field
$\delta$${}\!{\mbox{\boldmath $A$}}$=$\!\vec{\,\Omega}_{\!\tiny{\mbox{\boldmath $R$}}}
\!\!{}^{(l)}(\!\mbox{\boldmath $r$})$, which for its part determines an artificial
correction to the quantizing magnetic field,
\vspace{-2.mm}
\begin{equation}\label{corr_to_B}
\delta{ B}\!{}_\perp\!(\mbox{\boldmath $R$})=
\!\mbox{\boldmath $\nabla$}\!{}_{{\scriptsize\mbox{\boldmath $r$}}}
{}\!\!\times\!\!\vec{\,\Omega}{}_{\tiny{\!\mbox{\boldmath $R$}}}\!\!\!{}^{(z)}
(\mbox{\boldmath $r$}){}\!\left.\right|_{\mbox{\small{\boldmath $r$}}\to 0}
\equiv{}\!\!\Rot\!
{\vec{\,\Omega}}{}_{\!\tiny{\mbox{\boldmath $R$}}}\!\!\!{}^{(z)}, \vspace{-2.mm}
\end{equation}
directed also
perpendicular to the $\{{\hat X},
{\hat Y}\}$ plane. The adjusted
magnetic length becomes
${{\tilde l}_B}^{-1}\!=l_B^{-1}+{}l_B$${}\Rot\!\!
\vec{\,\Omega}\!$${}_{\tiny{\mbox{\boldmath $R$}}}$${}\!{}^{\!(z)}{}\!\!/2$, which
influences the ``compactness'' of the one-electron wave function
and, hence, the number of magnetic flux quanta per domain. The
latter is changed by $\Delta\!{}^2\!\mbox{\boldmath $R$}\cdot\delta B\!{}_\perp\!/2\pi$, where
$\Delta\!{}^2\!\mbox{\boldmath  $R$}\!\equiv\!\Delta X\Delta Y\equiv
\int_{{\mbox{\scriptsize{\boldmath $r$}}\in\,G\!{}_{\tiny{\mbox{\boldmath $R$}}}}}\!d{}^2\!\mbox{\boldmath  $r$}$
is the domain area.
Now by using a perturbation approach, we calculate the ground-state energy of the
$\textstyle{G}\!$${}_{\tiny{\mbox{\boldmath $R$}}}$ domain in the perpendicular
quantizing field \vspace{-2.mm}
\begin{equation}\label{tildeB}
{\widetilde B}\!=\!B_\perp\!+\delta B_\perp\,,\vspace{-2.mm}
\end{equation}
by counting this energy from the appropriate
value corresponding to the same domain in the same field ${\widetilde B}$, where,
however, the `$\vec{\,\Omega}^{(l)}\!\!{\hat \sigma}_l$'-terms
in the Hamiltonian are set equal to zero.
Certainly, with the perturbation approach, we have to hold equal the electron
numbers in the perturbed
and unperturbed systems. In the `global ground state', i.e. in the absence of
any spin rotation, the number of electrons
within the
$\textstyle{G}\!$${}_{\tiny{\mbox{\boldmath $R$}}}$ domain is equal to
${\cal N}$${}\!{}_{\scriptsize{\mbox{\boldmath $R$}}}$${=\nu\cdot
\Delta\!{}^2\!\mbox{\boldmath $R$}}\cdot B_\perp\!/2\pi$,
where $\nu$ is the factor equal to 1 or 2 depending on the type of quantum Hall
ferromagnet considered. In the state with spin rotation this
number is
changed by value\vspace{-2.mm}
\begin{equation}\label{tilde_q}
\delta\widetilde{q}={\nu\Delta\!{}^2\!\mbox{\boldmath $R$}}\cdot{}\Rot\!
\vec{\Omega}{}_{{}\!\tiny{\mbox{\boldmath $R$}}}^{(z)}\!/2\pi.\vspace{-1.5mm}
\end{equation}
The change of the
cyclotron energy compared to the `global ground state' is \vspace{-1.5mm}
\begin{equation}\label{cyclotron_corr}
\delta E_c^{(0,\nu)}=\displaystyle{\omega_c\frac{3\nu\!-2}{4\pi}}\,
\Delta\!{}^2\!\mbox{\boldmath $R$}\,\Rot\!\!
\vec{\,\Omega}{}_{\!\tiny{\mbox{\boldmath $R$}}}\!\!{}^{(z)},\vspace{-2.mm}
\end{equation}
where $\nu\!=\!1,$ or $\!2$. The contribution of the Coulomb interaction to the global ground-state energy is estimated as
$E^{(0,\nu)}_{\rm int}$$\sim$$\,{\mathscr E}_{\rm C}
{\cal N}$${}\!\!{}_{\scriptsize{\mbox{\boldmath $R$}}}$. Then the corrected
value, ${\widetilde E}^{(0,\nu)}_{\rm int}$ is proportional
to $1/{\tilde l}_B^3$, thus being
changed by\vspace{-2.mm}
\begin{equation}\label{Coul_corr}
{\delta E}^{(0,\nu)}_{\rm int}(\!\mbox{\boldmath $R$})\!=
3E^{(0,\nu)}_{\rm int}(\!\mbox{\boldmath $R$})l_B^2{}\Rot\!\!
\vec{\,\Omega}{}_{\!\tiny{\mbox{\boldmath $R$}}}\!\!\!{}^{(z)}\!/2 ,\vspace{-1.mm}
\end{equation}
as compared to the $E^{(0,\nu)}_{\rm int}(\!\mbox{\boldmath $R$})$ value.
The estimation of the $E^{(0,\nu)}_{\rm int}$ energy can be performed using the
Hartree-Fock formula,\cite{foot3}\vspace{-2.mm}
\begin{equation}\label{Coul_ground}
E^{(0,\nu)}_{\rm int}\!(\!\mbox{\boldmath $R$})
={}_{\mbox{\tiny\boldmath $R$}}\!\langle\nu, 0|\hat{\cal H}_{\mbox{\scriptsize int}}
(\mbox{\boldmath $R$})
|0,\nu\rangle\!{}_{\mbox{\tiny\boldmath $R$}}.\vspace{-1.mm}
\end{equation}
See Ref. \onlinecite{foot3} for details. $|0,\nu\rangle\!{}_{\mbox{\tiny\boldmath $R$}}$ in Eq. \ref{Coul_ground} means the `global'
ground state of the
$\textstyle{G}\!$${}_{\tiny{\mbox{\boldmath $R$}}}$ domain of
the $\nu=1,\,2$ quantum Hall ferromagnet.

\vspace{-2.mm}
\subsection{Perturbation theory results}
\vspace{-2.mm}

We have obtained corrections \ref{cyclotron_corr} and
\ref{Coul_corr} associated only with the renormalization of the effective magnitude
of the quantizing magnetic field \ref{tildeB}. Now we calculate corrections
to the ground state energy of the
$\textstyle{G}\!$${}_{\tiny{\mbox{\boldmath $R$}}}$
domain determined directly by the action of the perturbation operator\vspace{-2.mm}
$$
\widehat{V}_{\Omega}(\mbox{\boldmath $R$})\!=\!\displaystyle{\frac{1}{2m_e^*}
\int\limits_{G\!{}_{\tiny{\mbox{\boldmath $R$}}}}\!\!\!d^2\!\mbox{\boldmath $r$}
 \vec{\, \chi}{}\,^\dag\!\!\!{}_{\!\tiny{\mbox{\boldmath $R$}}}
 (\mbox{\boldmath $r$})\!\left\{\,\vphantom{
  \sum_l[\vec{\,\Omega}_{\!\tiny{\mbox{\boldmath $R$}}}\!\!{}^{(l)}]^2
\!(\mbox{\boldmath $r$})}
  \left[\vphantom{{\sum[\vec{\,\Omega}_{\!\tiny{\mbox{\boldmath $R$}}}\!\!{}^{(l)}]^2
\!(\mbox{\boldmath $r$})}}-i\mbox{\boldmath $\nabla$}\!{}_{{\scriptsize\mbox{\boldmath $r$}}}
  \!+\!{\mbox{\boldmath $A$}}_{\!\tiny{\mbox{\boldmath $R$}}}
  (\mbox{\boldmath $r$}\!)\qquad\qquad{}
\right.\right.}{}\vspace{-8.mm}
$$
\begin{equation}\label{Vomega}
+\!\!\!
{\sum_{l=x,y,z}}\!\!\!\vec{\,\Omega}_{\!\tiny{\mbox{\boldmath $R$}}}\!\!{}^{(l)}
\!(\mbox{\boldmath $r$}\!)
\hat{\sigma}_l\left.\vphantom{\sum[\vec{\,\Omega}_{\!\tiny{\mbox{\boldmath $R$}}}\!\!{}^{(l)}]^2
\!(\mbox{\boldmath $r$})}\!\right]^2
\!\!\!-\!\left[\vphantom{\sum[\vec{\,\Omega}_{\!\tiny{\mbox{\boldmath $R$}}}\!\!{}^{(l)}]^2
\!(\mbox{\boldmath $r$})}\!-i\mbox{\boldmath $\nabla$}
\!{}_{{\scriptsize\mbox{\boldmath $r$}}}
 \!+\!\!{\mbox{\boldmath $A$}}_{\!\tiny{\mbox{\boldmath $R$}}}
  (\mbox{\boldmath $r$}\!)\right]^2\!\left.\vphantom{
  \sum_l[\vec{\,\Omega}_{\!\tiny{\mbox{\boldmath $R$}}}\!\!{}^{(l)}]^2
\!(\mbox{\boldmath $r$})}\!\right\}\!
  \vec{\, \chi}\!{}_{\tiny{\mbox{\boldmath $R$}}}(
  \mbox{\boldmath $r$})\!\!\!\!{}\vspace{-2.5mm}
\end{equation}
on the
ground state $|0,\nu\rangle\!{}_{\mbox{\tiny\boldmath $R$}}$.
Opening
the square brackets in this expression and using ordinary manipulations,
we arrive at \vspace{-2.mm}
\begin{equation}\label{U+W}
{\widehat V}_{\Omega}\approx
   {\widehat {\cal U}}+
   {\widehat {\cal W}}\,\vspace{-2.mm}
\end{equation}
(see Ref. \onlinecite{di02}), where operator ${\widehat {\cal U}}$ is spinless,
whereas ${\widehat {\cal W}}$ leads to a
 spin flip (i.e.
to changes $S_z\to S_z\!-\!1$ and $S\to S\!-\!1$). Let ${\hat c}_{np}$ and ${\hat c}_{\overline{n}p}$
be operators annihilating electrons on the $n$-th Landau level in the spin
`up' and `down' states respectively ($p$ is the notation for orbital quantum
states within a Landau level).
Then \vspace{-3.mm}
\begin{equation}\label{U}
\begin{array}{l}
\displaystyle{{\widehat {\cal U}}\!=\!\frac{l_B^2{\omega}_c}{2}\!}\left[\sum_{l=x,y,z}
\!\left({\vec \Omega}^{(l)}\right)^2\right.\!{\hat{\cal N}}\vspace{-1.mm}\\
\quad\displaystyle{
+\Rot\!{\vec\Omega}^{(z)}}\!\!\sum_n\!\displaystyle{\left.\vphantom{\sum_{l}
\!\left({\vec \Omega}^{(l)}\right)^2\!\sum_n{\hat{N}}_n}\!\left(2n+1\right){\hat N}_n\right]}
\!+l_B\omega_c\Omega_-^{(z)}{\hat K}^\dag
\end{array}\vspace{-4.mm}
\end{equation}
and \vspace{-1.mm}
\begin{equation}\label{W}
{}\!{}\!{}\!{}{\widehat {\cal W}}\!\!=\!l_B{\omega}_c\!\sum_n\!\sqrt{n+1}\!\left(\!\Omega_+^+
  {\hat{\cal Q}}^\dag_{n+1\,\overline{n}}\!+\!\Omega_-^+
  {\hat{\cal Q}}^\dag_{n\,\overline{n+1}}
  \right)\!,  \vspace{-2.5mm}
\end{equation}
where
\begin{equation}\label{Omega+/-}
{}\!{}\!{}\!{}\!{}\Omega^{(l)}_{\pm}\!\!=\!\mp\!\frac{i}{\sqrt{2}}
  \left[\Omega_x^{(l)}\!\pm i\Omega^{(l)}_y\right]\!,\,
\Omega_{\mu}^{\pm}\!\!=
  \!\left[\Omega^{(x)}_{\mu}\!\!\pm\!i\Omega^{(y)}_{\mu}\right]\!;\vspace{-5.mm}
\end{equation}
$$
\begin{array}{l}
{{\hat N}_{n}=\sum_p\left({\hat c}^\dag_{np}{\hat c}_{np}\!+
{\hat c}^\dag_{\overline{n}p}{\hat c}_{\overline{n}p}\right)\,,\qquad
\hat{\cal N}\!=\!\sum_n{\hat N}_{n}}\,,\qquad\qquad\vspace{2.mm}\\
\!{{\hat K}^\dag\!\!=\!
\sum_{np}\!\sqrt{n+1}
  {\hat c}^\dag_{{n\!+\!1}p}{\hat c}_{np}\,,}\:\:
\mbox{and}\:\:\,{{\hat{\cal Q}}^\dag_{n\overline{m}}\!=
  \!\sum_{p}{\hat c}^\dag_{\overline{m}p}{\hat c}_{np}}\,.\vspace{0.mm}
\end{array}
$$
In Eqs. \ref{U} and \ref{W} we keep only the operators
that give a nonzero result when acting on fully spin-polarized ground
state
$|0,\nu\rangle\!{}_{\mbox{\tiny\boldmath $R$}}$. We omit also subscript
...${}_{\mbox{\scriptsize\boldmath $R$}}$ in these equations and everywhere further,
not forgeting that Eqs. \ref{U} and \ref{W} apply only to
the $\textstyle{G}\!$${}_{\tiny{\mbox{\boldmath $R$}}}$ domain.
The sign of the approximate equality in Eq. \ref{U+W} means that we
omitted the terms leading to
corrections to energy of a higher order than the second one in the terms
of the gradients of the Euler rotation angles.

The operator of the total number of particles $\hat{\cal N}$ is certainly
diagonal for our system with a fixed number of electrons. $\hat{K}^\dag$ is
the raising
ladder operator,\cite{ko61} and, when acting on any eigen state of our system,
it always results in an eigen state with energy higher by cyclotron one
$\omega_c$. This property of the operator $\hat{K}^\dag$, as well as the diagonality of the operator $\hat{\cal N}$, are general and hold irrespective of the chosen model. $\hat{K}^\dag$ is contributing to the second order. Operator $\hat{N}_n$
(corresponding to the number of electrons on the Landau level $n$), acting on our fully polarized
ground state $|0,\nu\rangle$, gives $N_\phi|0,\nu\rangle\delta_{n,0}$ if
$\nu\!=\!1$, or
$N_\phi|0,\nu\rangle\left(\delta_{n,0}+\delta_{n,1}\right)$ if $\nu\!=\!2$; where\vspace{-2mm}
\begin{equation}\label{Nphi}
N_\phi=\!\Delta\!{}^2\!\mbox{\boldmath $R$}/2\pi l_B^2\vspace{-2mm}
\end{equation}
is the number of
the maganetic flux quanta in the
$\textstyle{G}\!$${}_{\tiny{\mbox{\boldmath $R$}}}$ domain.
As a result, we obtain the perturbation theory correction
determined by operator \ref{U} at filling factors $\nu=1$ or $2\,$:\vspace{-2.mm}
\begin{equation}\label{corrU}
{}\!{}\!{}\!\delta E\!{}^{(\!\nu\!)}_{\cal U}\!\!=\!\omega_cN_\phi l_B^2\!\!
\left[\!\frac{\nu}{2}\!\!\!\sum_{{}\,\,l=x,y}
\!\!\!\left(\!{\vec \Omega}^{(l)}\!\right)\!\!{\vphantom{\left({\vec \Omega}^{(l)}\!\right)}}^2\!\!
 \!+\!\left(\!\frac{3\nu}{2}\!-\!1\!\right)\Rot{{}\!\vec\Omega}^{(z)}\!{}\!\right]\!\!.\!\!{}
\vspace{-2.mm}
\end{equation}

The perturbative correction to the ground-state energy determined by
spin-cyclotron operator
\ref{W}, appears only in the second order.  This operator, unlike ${\hat K}^\dag$,
does not commute with interaction Hamiltonian \ref{CoulR}, and, hence, leads to a
significant mixing of Landau levels. Generally, to find the desired correction, we
should consider as a basic set all kinds of spin-flip--orbital excitations mixing
various Landau levels. Virtual
transitions from the ground state to these excitations determine the denominators in the
formula for the ${\cal W}$-correction. If $r_s\gtrsim 1$, the
denominators are of the order of
${\mathscr E}_{\rm C}$, and thus the ${\cal W}$-corrections vanish at
$r_s\gg 1$.

Among all the possible modes of various collective spin-flip states that could
form a complete
basic set for calculating the ${\cal W}$-correction,  there are certainly single-mode
(single-exciton) states. Excitations in QH systems can  be studied within the
framework of this single-exciton basis, and sometimes at integer filling factors such an approach even yields an asymptotically exact result to the first order in small
$r_s$.\cite{by81,pi92} Now, studying a system with an arbitrary value of $r_s$,
we, as in Ref. \onlinecite{di20}, use the single-exciton basis as a model to describe
spin-flip states that are relevant for calculating the ${\cal W}$-correction. Then
the basic set consists only of orthogonal single-`excitonic' states with the
${\vec q}\!=\!0$
wave vectors: \vspace{-1.5mm}
$$|n\overline{m},\nu\rangle\!=\!N_\phi^{-1/2}{\hat Q}^\dag_{n\overline{m}}
|0,\nu\rangle,\vspace{-1.5mm}
$$
where $n=0\,\,\mbox{or}\,\,1$
and $m\!\neq\!n$. Specifically, in the $\nu\!=\!1$ case, the single-mode basis
relevant for calculating the ${\cal W}$-correction is presented by the only state
$|0\overline{1},1\rangle$, since in this case the quantum mixing
appearing due to the Coulomb correlation
between
 state $\widehat{\cal W}|0,1\rangle$ and spin-flip states
 $|n\overline{m},1\rangle$ (i.e.
 $\propto\langle 1,\overline{m}n|\hat{\cal H}_{\rm int}
 |\widehat{\cal W}|0,1\rangle$),
does not vanish only if $n\!=\!0$ and $m\!=\!1$. The energy of this
excitation, counted from the level of the ground state, is
$\;\omega_c \!+
\epsilon_{\rm Z}\!+{\cal E}_{0 \overline{1}}\,,\,$ where\vspace{-1.5mm}
{}\!{}\!{}\!\begin{equation}\label{E01}
{}\!{\cal E}_{0 \overline{1}}=\!\langle 1,\!\overline{1}0|\hat{\cal H}_{\rm int}
|0\overline{1},\!1\rangle
\displaystyle{=\!\!{}\!\int_0^{\infty}\!\!\!\!\!dp(p^3\!/2){\tilde V}(p)
e^{{}\!-p^2\!/2}}\! \vspace{-2.mm}
\end{equation}
(see Ref. \onlinecite{di20}), where\vspace{-2.mm}
\begin{equation}\label{tildeV}
{\tilde V}(p)
\!\equiv\!\int\!V(\mbox{\boldmath{$r$}})
e^{-i{\mbox{\small\boldmath{$p$}}\mbox{\small\boldmath{$r$}}\!/\mbox{\small $l_B$} }}d^2\!\mbox{\boldmath{$r$}}\!/2\pi l_B^2.\vspace{-2.mm}
\end{equation}
It corresponds to
the found $0\!\to\!\overline{1}$ spin-flip excitation with a zero wave vector. So, the
${\cal W}$-correction,
$\delta E\!{}^{(\!\nu\!)}_{\cal W}$, is obtained in the framework of the single-exciton
basic set at $\nu\!=\!1\,$:\vspace{-2.mm}
\begin{equation}\label{corr1W}
\delta E\!{}^{(1)}_{\cal W}\approx-N_\phi(l_B\omega_c)^2
\Omega_+^-\Omega_-^+\!/(\omega_c\!+\!{\cal E}_{0 \overline{1}})\vspace{-2.mm}
\end{equation}
(the value of $\epsilon_{\rm Z}$ is neglected compared to
$\omega_c\!+\!{\cal E}_{0 \overline{1}}$). The final results describing the
skyrmion/antiskyrmion excitations at the $\nu\!=\!1$ filling factor,
obtained by means of the present approach, are given in Ref. \onlinecite{di02}.

The spin-flip eigen-states in a $\nu\!=\!2$ quantum Hall ferromagnet, within the
framework of the single-exciton set $|n\,\overline{m},2\rangle$, were studied in
the study of Ref. \onlinecite{di20}. The ${\cal W}$-correction
to the ground state $|0,2\rangle$ is found from the relevant basic set  consisting of three
states: \vspace{-5.mm}
$$\!{}
|1\overline{0},2\rangle\quad\mbox{and}\quad|\pm,2\rangle=
\displaystyle{\frac{A_\pm|0\overline{1},2\rangle+
B_\pm|1\overline{2},2\rangle}{({1+A_+^2})^{1\!/2}}}.\vspace{-2.5mm}
$$
\vspace{1.mm}The notations used are:

\noindent$A_+\!=\!B_-\equiv(a\!-\!b)/d{+\sqrt{(a\!-\!b)^2\!/h^2+1}},\vspace{.5mm}\,$

\noindent and $A_-\!=\!-B_+\!=1$, where\vspace{-2.5mm}
\begin{equation}\label{a}
{}\!\!\!{}\!{}a=\!\displaystyle{\int_0^\infty\!\!\!
\tilde{V}(p)e^{-p^2\!/2}p^3dp}\,,\qquad\qquad\qquad\qquad\vspace{-5.mm}
\end{equation}
\begin{equation}\label{b}
b=\!\!{}
\displaystyle{\int_0^\infty\!\!\!{\tilde V}(p)e^{-p^2\!/2}\left(1+{p^4}\!/
{16}-3{p^2}\!/{8}\right)p^3dp}\,,\vspace{-2.mm}
\end{equation}
and\vspace{-2.mm}
\begin{equation}\label{h}
{}\!\!\!{}\!{}\!h\!=\!
\displaystyle{\frac{1}{\sqrt{2}}\left|\int_0^\infty\!\!\!{\tilde V}(p)e^{-p^2\!/2}
\left(2-p^2\!/2\right)p^{3}dp\right|\,.}\vspace{-1.mm}\end{equation}
The energies of these excitations are equal to
$\epsilon_{\rm Z}-\omega_c\!+{\cal E}_{1\overline{0}}\,$ and
$\epsilon_{\rm Z}+\omega_c+{\cal E}_\pm\,$ respectively, where\vspace{-2.mm}
\begin{equation}\label{energies}
{}\!{}\!{}\begin{array}{l}
{\cal E}_{1\overline{0}}=\displaystyle{\int_0^\infty\!\!{\tilde V}(p)e^{-p^2\!/2}p^5dp/4}\quad\mbox{and}\qquad\qquad\vspace{1.mm}\\
\displaystyle{{\cal E}_\pm=\left({a+b\pm\sqrt{{(a-b)^2}+h^2}}\right)\!/2}\,.\qquad
\end{array}\vspace{-2.mm}
\end{equation}
Thus we obtain the correction determined by
operator \ref{W}:
\vspace{-2.5mm}
\begin{equation}\label{corr2W}
{}\!{}\begin{array}{l}
\delta E\!{}^{(2)}_{\cal W}\!\approx\!N_\phi(l_B\omega_c)^2\displaystyle{\left\{\frac{
\Omega_+^+\Omega_-^-}{\omega_c\!-\!{\cal E}_{1 \overline{0}}}\right.}\vspace{1.mm}\\\displaystyle{-
\frac{\Omega_+^-\Omega_-^+}{1\!+\!A_+^2}\left.\!\!\left[\!\frac{(A_+\!-\!\sqrt{2})^2}
{\omega_c\!+\!
{\cal E}_{+}}+\frac{(1\!+\!\sqrt{2}A_+)^2}{\omega_c\!+\!
{\cal E}_{-}}\!\right]\right\}},
\vspace{-2.mm}\end{array}
\end{equation}
again neglecting the $\epsilon_{\rm Z}$ value compared to
$\max{\!(\omega_c,\, {\mathscr E}_{\rm C})}$.

\subsection{Energy of the skyrmion-antiskyrmion pair excitation}

The meaning of formulae \ref{corr_to_B}, and \ref{tilde_q} -- \ref{Coul_corr}
is revealed by a very important feature of the value
$\Rot\!\vec{\,\Omega}$${}{}_{\!\tiny{\mbox{\boldmath $R$}}}$${}\!\!\!{}^{(\!z)}$.
Specifically, if we use the expression for ${\vec \Omega}^{(z)}$
through Euler angles $\alpha(\mbox{\boldmath $R$})$ and
$\beta(\mbox{\boldmath $R$})$ [see Refs. \onlinecite{io99} and \onlinecite{di02},  and
Eqs. \ref{top_density} and \ref{rot_Omegaz} in Appendices below], it turns
out that\vspace{-3.mm}
\begin{equation}\label{Rot=rho}
\Rot\Omega{}_{\tiny{\mbox{\boldmath $R$}}}\!\!\!\!{}^{(\!z)}\equiv
-2\pi\rho_{\mbox{\tiny T}}(\mbox{\boldmath $R$}),\vspace{-1.mm}
\end{equation}
where the topological density is given by equation \ref{top_density}. According
to Eq. \ref{tilde_q}, the electron density at point $\mbox{\boldmath $R$}$
is changed by
\begin{equation}\label{rho_s}
  \delta\rho_s=-\nu\rho_{\mbox{\tiny T}}(\mbox{\boldmath $R$}).
\end{equation}
That is, the actual electric charge attributed
to the studied state is topological charge
$q_{\mbox{\tiny T}}\!=\!\int\rho_{\mbox{\tiny T}}(\mbox{\boldmath $R$})d^2\mbox{\boldmath $R$}$,
multiplied by integer factor $\nu$.  In the case of the
$\nu\!=\!2$ ferromagnet, the skyrmion ($q_{\mbox{\tiny T}}\!=\!-1$) and
antiskyrmion ($q_{\mbox{\tiny T}}\!=\!1$) have respectively negative and positive electric
charges equal in magnitude to two elementary charges.

The meaning of the correction given by Eq.
\ref{cyclotron_corr} becomes trivial: after summing over all domains
$\textstyle{G}\!$${}_{\tiny{\mbox{\boldmath $R$}}}$ [see Eq. \ref{dR}],
this represents a change in the single-particle orbital
energy due to the resulting  electron excess or deficiency in the system in question.
In fact, at a fixed total number
of particles in the system, the excess and deficiency cancel each other, and
the total correction to the cyclotron energy determined by Eq. \ref{cyclotron_corr}
vanishes. At the same time, the energy gap for neutral spin excitation acquires
a clear physical meaning. In our problem, this is excitation of a
skyrmion-antiskirmion pair with oppositely charged components separated by
a large distance, so that the interaction among them can be neglected.
When performing
summation/integration over $\mbox{\boldmath $R$}$ [see Eq. \ref{dR}] of various
contributions \ref{Coul_corr},
\ref{corrU}, \ref{corr1W} and \ref{corr2W} to the total skyrmion-antiskirmion energy,
we simply omit the terms proportional to
$\Rot\!\!
\vec{\,\Omega}\!$${}_{\tiny{\mbox{\boldmath $R$}}}$${}\!\!{}^{\!(\!z)}$, as canceling
each other.

Along with Eq. \ref{Rot=rho}, there are other identities
relating the
$\vec{\,\Omega}\!$${}_{\tiny{\mbox{\boldmath $R$}}}$${}\!\!{}^{\!(\!z)}$
components with
the field ${\vec n}(\mbox{\boldmath $R$})$ (see
Appendix B), and thus allowing presentation of the results in terms
of spatial derivatives of vector ${\vec n}$. At the
$\nu\!=\!2$ filling factor (the case considered in details) and,
in Eqs.\ref{corrU}, and \ref{corr2W} keeping only the terms
that do not contain $\Rot\!\!
\vec{\,\Omega}\!$${}_{\tiny{\mbox{\boldmath $R$}}}$${}\!{}^{\!\!(\!z)}$,
via summing/integrating
over $\mbox{\boldmath $R$}$, we obtain the contribution to the gap of creation
of the skyrmion-antiskyrmion pair:\vspace{-1.mm}
\begin{equation}\label{gap_D}
\begin{array}{l}
{}\displaystyle{{}\:D=2\sum_{{\mbox{\scriptsize\boldmath $R$}}}\!\left[\delta E^{(2)}_U+
\!\delta E^{(2)}_W\right]}\vspace{-1.mm}\\
\qquad\qquad\quad\displaystyle{ =\frac{J}{2}\!\int\!\!
\left[\left(\partial_X\!{\vec{\, n}}\right)^2\!+\!
  \left(\partial_Y\!{\vec{\, n}}\right)^2\right]\!d^2\!\mbox{\boldmath $R$}},
\end{array}\vspace{-4.mm}
 \end{equation}
 where\vspace{-2.mm}
 \begin{equation}\label{gapJ}
 \begin{array}{l}
J\!=\!\displaystyle{\frac{\omega_c^2}{4\pi}\!\left[\vphantom{\displaystyle{-
\frac{(A_+\!-\!\sqrt{2}\!)^2}{(1\!+\!B_+^2)(\omega_c\!+\!\!
{\cal E}_{+})}
\!-\!\left. \frac{(A_+\!-\!\sqrt{2}\!)^2}{(1\!+\!B_+^2)(\omega_c\!+\!
{\cal E}_{-})}\!\right]}}\frac{2}{\omega_c}\!+\!
\frac{1}{\omega_c\!-\!{\cal E}_{1 \overline{0}}}\right.}\vspace{1.mm}\\
\displaystyle{\;-
\frac{(A_+\!-\!\sqrt{2})^2}{(1\!+\!A_+^2)(\omega_c\!+\!\!
{\cal E}_{+})}
-\left. \frac{(1\!+\!\sqrt{2}A_+)^2}{(1\!+\!A_+^2)(\omega_c\!+\!
{\cal E}_{-})}\!\right]}\!\!
\end{array}\vspace{-1.mm}
\end{equation}
[see Eqs. \ref{corrU} and \ref{corr2W}, and identities \ref{rot_Omegaz}-
\ref{Omegan} in
Appendix B]. According to the main result of the NL$\sigma$ model [see Eq. \ref{minE} in
Appendix A], the lowest non-trivial (not equal to zero) minimum of this value is
achieved when the integral in Eq. \ref{gap_D} is equal to
$8\pi$, i.e. when the topological charge is $|q_{\mbox{\tiny T}}|=1$.

Note that the calculated gap,\vspace{-1.mm}
\begin{equation}\label{DJ}
D\!=\!4\pi J,\vspace{-1.mm}
\end{equation}
appears only owing to the electron-electron
correlations. It vanishes if we equate to zero the values
${\cal E}_{1 \overline{0}}$ and ${\cal E}_{\pm}$ proportional to the Coulomb vertex.

When $r_s\!\gg\!1$, we obtain, as predicted above, $D$ approaching $2\omega_c$, and
thus weakly depending on the interaction. (At unit filling
$\nu$, we have the $r_s\!\gg\!1$ value twice smaller: $D\approx
\omega_c$.\cite{di02})
This result is valid at the zero value of the Zeeman gap and, therefore,
in the absence of anything that
somehow limits the $\Lambda$ scale.
However, some dependence on the interaction
appears with finite $\epsilon_{\rm Z}$, which determines the real
value of $\Lambda$ (large compared to $l_B$, yet finite).
The situation is similar to that
of cyclotron resonance frequency, for which there is no dependence on the
electron-electron interaction in a translationally invariant system,\cite{ko61}
although it appears as soon as this invariance is broken.

\section{ Discussion of
the results}

So, the calculated value $D$ [Exs. \ref{gap_D}-\ref{DJ}] represents the
exchange contribution and an essential part of the Coulomb contribution
to the transport gap. Both constitute a comprehensive result for the transport gap
in the asymptotic limit $\epsilon_{\rm Z}\!/\!{\mathscr E}_{\rm C}\to 0$ (condition that determines the limit
$\Lambda\!\to\!\infty$.\cite{by98} At
fixed filing factor $\nu\!=\!2$, the dependence of $D$  on the magnetic field is shown
in Fig. 1. The calculation was performed for a specific material parameter
corresponding to a ZnO/MgZnO heterostructure ($m_e^*=0.3m_e$ and $\kappa=8.5$).
Besides, the renormalized $e$-$e$ interaction vertex is
chosen in the form  ${\tilde V}(p)=e^2/\kappa l_Bp(1\!+\!dp)$, which was used
earlier.\cite{di20} The figure shows also transport gap $\Delta_1$ related to another
type of excitation in a $\nu\!=\!2$ QH ferromagnet, namely, excitation of an
electron--exchange-hole pair.\cite{foot}
It is seen that: (i) the $D(B)$ value is appreciably smaller than the electron--exchange-hole gap
$\Delta_1$ calculated within the same approach; (ii) $D(B)$ represents the non-monotonic function of
$B$, and has a maximum at $B\gtrsim 5\,$T; and (iii) at some $B>8\,$T, depending on the effective
quantum well width $d$ (presented in $l_B$ units),
the calculated $D(B)$ vanishes, which, in fact, points to the feasibility of a Stoner transition to the paramagnetic phase.
\vspace{-5.mm}
\begin{figure}[h]
\begin{center}
\hspace*{-15.mm}
\vspace{-2.mm}
\includegraphics[width=.65\textwidth]{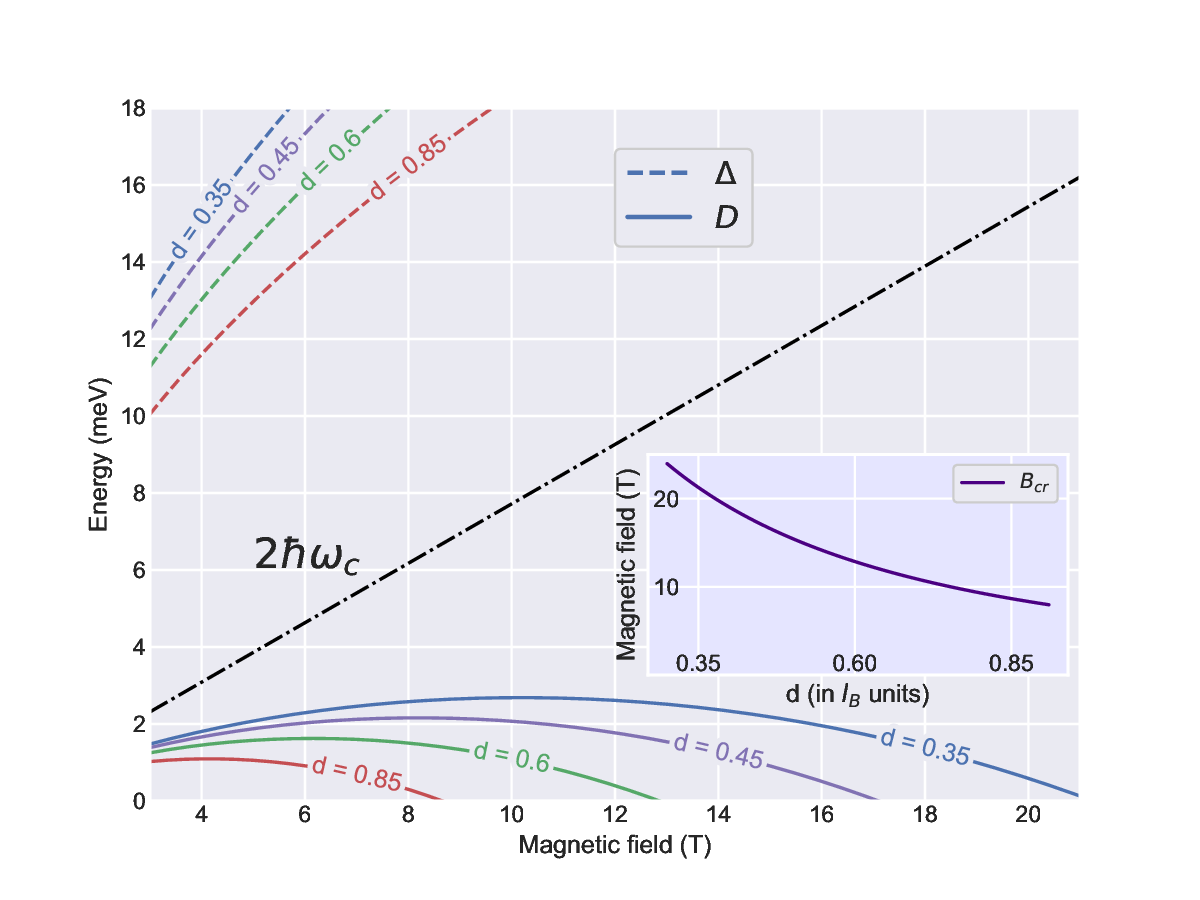}
\end{center}
\vspace{-8.mm}
\caption{Solid lines: the calculated skyrmion-antiskyrmion excitation gap $D(B)$ [Eqs. \ref{gapJ} and \ref{DJ}; material parameters are $m_e^*=0.3m_e$ and $\kappa=8.5$] at different quantum well effective width $d$ parametrizing the 2D Fourier component ${\tilde V}(p)$ of the Coulomb vertex [see equation (2.29) and the specific expression for ${\tilde V}$ given in the text of section III]. Value $d$ is given in units of the magnetic length. Dash lines: the electron-exchange-hole gap $\Delta$ (see Ref. \onlinecite{foot}) calculated in the framework of the same approach/model. On the inset: the critical magnetic field $B_{cr}$, where $D$ vanishes [$D(B_{cr})=0$], is shown as function of parameter $d$.}
\end{figure}
\vspace{-3.mm}

We emphasize that the study presented is purely theoretical.
However, it is worth noting that the actual situation is as follows:
The conditions
under which the $D$ value gives the main contribution to the creation energy of
the skyrmion-antiskyrmion pair, and thereby
to the transport gap, are hardly met in the QHSs currently
investigated in experiments.\cite{ma14,va17}
Indeed, in order to ignore the change of the Zeeman energy,\vspace{-2.5mm}
\begin{equation}\label{Zeeman}
\delta E_{\rm Z}=\frac{\nu\,\epsilon_{\rm Z}}{2}\!\int\!\! d^2\!\mbox{{\boldmath $R$}}
\,\left[1-n_z(\!\mbox{{\boldmath $R$}})\right],\vspace{-1.5mm}
\end{equation}
and to neglect the Coulomb
(Hartree) interaction between different charged
domains ${G}$$\!{}_{\footnotesize{\mbox{\boldmath $R$}}_i}$ (determined only
by the inter-domain repulsion at
distances $|\mbox{\boldmath $R$}_i-\mbox{\boldmath $R$}_j|\gg l_B$),\vspace{-1.mm}
\begin{equation}\label{Hartree}
V_{\rm H}=\displaystyle{\!\frac{\nu^2e^2}{2\kappa}\!
\int\!\!\!\int\!\!\frac{d^2\!\!\mbox{{\boldmath $R$}}\,
  d^2\!\!\mbox{{\boldmath $R$}}'}{|\!\mbox{{\boldmath $R$}}\!-\!\!\mbox{{\boldmath $R$}}'|}
\rho_{\mbox{\tiny T}}\!(\!\mbox{{\boldmath $R$}})\rho_{\mbox{\tiny T}}\!
(\!\mbox{{\boldmath $R$}}')}\vspace{-1.mm}
\end{equation}
[see Eqs. \ref{Coul} and \ref{CoulR}, and also cf., e.g., Refs. \onlinecite{so93}
and \onlinecite{by98}], it is necessary that the ratio $\epsilon_{\rm Z}/(e^2/\kappa l_B)$
be not simply small, but its smallness must be such that the logarithm
$\ln{\!(e^2\!/\kappa l_B\epsilon_{\rm Z})}$ is large.\cite{by98} (See also Appendix A below.) An estimate, that is easy to make in the same way as it was done earlier in the works devoted to the $\nu\!=\!1$ ferromagnet,\cite{by98} leads to the conclusion: `classical'
corrections, given by Eqs. \ref{Zeeman} and \ref{Hartree}, become essentially smaller than the value \ref{gapJ}--\ref{DJ} only in the situation where
$\epsilon_{\rm Z}/(e^2\!/\kappa l_B)\!<\!0.001$. (So then $\Lambda$ turns out to be well larger than $l_B$, indeed.) Whereas, even for GaAs/AlGaAs
2D structures the characteristic value is
$\epsilon_{\rm Z}/(e^2\!/\kappa l_B)\!\backsimeq\!0.01$,
and for ZnO/MgZnO quantum wells we get it $\gtrsim\!0.03$.

Apparently, there are certain techniques that can reduce the value of $\epsilon_{\rm Z}$ experimentally; that is, to reduce
effectively the Land$\acute{\mbox{e}}$ factor $g$ in actual experiments (see, for
instance,
Ref. \onlinecite{ma96}). Then the calculated value $D$ can correspond to the
energy gap of creation of charge carriers, skyrmions and antiskyrmions, responsible
for Ohmic transport in ZnO/MgZnO quantum heterostructures. More probable
(even without artificial suppression of the Land$\acute{\mbox{e}}$ factor)
is appearance
of a spin-charge texture in the $\nu\!=\!2$ QH ferromagnet ground state
near the critical field $B_{\rm cr}$, corresponding to the Stoner transition.
This texture should be characterized by a local spin change with amplitude
$\delta S\!>\!1$ and the correlation length $\Lambda\!>\!l_B$.

In conclusion, we note that only the \ref{gapJ}--\ref{DJ} result,
of microscopic calculation, where the exchange interaction is appropriately taken into account,
can predict the Stoner
transition to a paramagnet phase. Neither the Zeeman energy \ref{Zeeman} nor the Hartree energy
\ref{Hartree} (both smoothly growing with the magnetic field) give any
grounds for the possibility of such a phase transformation.

The authors are grateful to A.V. Shchepetilnikov and A.B. Van'kov for useful discussion, and the Russian Science Foundation
(Grant No. 22-12-00257) for support.
\vspace{-0.mm}

\vspace{10.mm}
\appendix

\vspace{-2.mm}
\section{The \mbox{{\boldmath $O(3)$}} nonlinear
sigma model}

For convenience, we review  some relevant results of the classical field theory.
In the framework of the nonlinear
sigma (NL$\sigma$) model$\,${}\cite{be75,ra89}, the density of `gradient' energy
$e(\mbox{{\boldmath $R$}})$
and topological
density $\rho_{\mbox{\tiny T}}(\mbox{{\boldmath $R$}})$ are given by expressions\vspace{-1.mm}
\begin{equation}\label{sigma}
\begin{array}{l}
\displaystyle{e(\mbox{{\boldmath $R$}})\!=\!\frac{J}{2}
\left[\left(\partial_X\!{\vec{\, n}}\right)^2\!+\!
  \left(\partial_Y\!{\vec{\, n}}\right)^2\right]}\qquad\qquad\qquad\vspace{.5mm}\\
  \qquad\qquad\qquad\displaystyle{\equiv\frac{J}{2}\!\!\left[\left(\!\!\mbox{
{\boldmath $\nabla$}}{\beta}\right)^2\!\!
+\!\sin^2\!\!{\beta}\left(\!\mbox{{\boldmath $\nabla$}}
{\alpha}\right)^2\right]}\,;
\end{array}
\vspace{-2.mm}
\end{equation}
and\vspace{-2.mm}
\begin{equation}\label{top_density}
{}\!\!\begin{array}{l}\displaystyle{\rho_{\mbox{\tiny T}}(\mbox{{\boldmath $R$}})=
(4\pi)^{-1}{\vec{\,n}}\cdot
\left(\partial_X\!
  {\vec{\,n}}\right)
  \times\left(\partial_Y\!{\vec{\,n}}\right)\qquad\qquad}\vspace{.5mm}\\
  \displaystyle{\quad\equiv\,(4\pi)^{-1}\sin\!\beta\cdot
  \left(\partial_X\beta\cdot
  \partial_Y\alpha-\partial_Y\beta\cdot
  \partial_X\alpha\right)},
  \end{array}\vspace{-1.mm}
\end{equation}
where ${\vec n}$ is the 3D unit vector \ref{vec_n}, $\mbox{{\boldmath $\nabla$}}
\equiv (\partial_X,\partial_Y)$, and $J$ is
the spin stiffness -- parameter undefined in the framework of the NL$\sigma$ model,
which, however can be found microscopically (see Sec. III).
Both values, $e(\mbox{{\boldmath $R$}})$ and $\rho_{\mbox{\tiny T}}(\mbox{{\boldmath $R$}})$,
are also invariant with respect to substitution \ref{phi}.

With the help of Eqs. \ref{vec_i} we find that $e(\mbox{{\boldmath $R$}})$ and
$\rho_{\mbox{\tiny T}}(\mbox{{\boldmath $R$}})$
are expressed in terms of $(n_X,n_Y\!,n_Z)$
as well as in terms of $(n_x,n_y,n_z)$  above, i.e. the equations of the NL$\sigma$ model
are invariant with respect to rotation by a constant
angle $\theta$. However, the limit values of $\vec{\, n}$ at $\mbox{{\boldmath $R$}}\!=\!0$ and
$R\!=\!\infty$ are transformed into
$n_X|_{R=0}\!=\!-\sin\theta,\:n_Y|_{R=0}\!=\!0,\:n_Z|_{R=0}\!=\!-\cos\theta$ and
$n_X|_{R=\infty}\!=\!\sin\theta,\:n_Y|_{R=\infty}\!=\!0,\:
n_Z|_{R=\infty}\!=\!\cos\theta$,
respectively.

The main features of the  NL$\sigma$ model are as follows:\cite{be75}
(i) since the continuous, suitably behaved, function $\vec{\, n}(\mbox{{\boldmath $R$}})$
implements the mapping of the $\{{\hat X},{\hat Y}\}$ plane onto a unit sphere
parametrized by angles
$\alpha$ and $\beta$, the topological charge $q_{\mbox{\tiny T}}[\vec{\, n}]\!=
\!\int\!\rho_{\mbox{\tiny T}}(\mbox{{\boldmath $R$}})d^2\!\mbox{{\boldmath $R$}}$
takes only integer-number values, either positive or
negative, depending on the function $\beta(\mbox{{\boldmath $R$}})$, running through values from $\left.\beta(\mbox{{\boldmath $R$}})\right|_{R\!\to\infty}\!\!=\! 0$  to $\beta(\mbox{{\boldmath $0$}})=m\pi$, where $m=\pm1,\pm2,...;\,$ (ii) the minima of the energy
$E[\vec{\, n}]\!=\!\int\! e(\mbox{{\boldmath $R$}})d^2\!\mbox{{\boldmath $R$}}\,$,
considered as a function of $\vec{\, n}(\mbox{{\boldmath $R$}})$, are determined by
the $q_{\mbox{\tiny T}}$ values,
\begin{equation}\label{minE}
  \min{E[\vec{\, n}]}=4\pi J|q_{\mbox{\tiny T}}|\,.
\end{equation}

It is known that
$w=\!\cot{\!(\beta\!/\!2)}e^{i\alpha}$ represents an
analytical functions of the variable ${\mathscr Z}=X+iY$.\cite{be75} This property and conditions
of the physically appropriate behavior of the ${\vec n}(\alpha, \beta)$ vector considered as
a function of
$\mbox{{\boldmath $R$}}$ enable to find an analytical form of $w({\mathscr Z})$.
In particular, if
${\vec n}(\mbox{{\boldmath $R$}})$ has no singularities
at finite $\mbox{{\boldmath $R$}}$, then in the simplest but non-trivial case (i.e. when $w$
is not equal to
a constant) the unit topological charge,
$q_{\mbox{\tiny T}}\!=\!\pm\!1$, corresponds to a field where $\,$\cite{be75}\vspace{-1.mm}
\begin{equation}\label{w_z}
w={\mathscr Z}/\Lambda .\vspace{-1.mm}
\end{equation}
Within the framework of the macroscopic approach used, the parameter $\Lambda$
controlling the size scale remains undetermined within the NL$\sigma$.
model. From Eq. \ref{w_z} it follows
that in this case \vspace{-2.mm}
\begin{equation}\label{vec_nRXY_}
\cos{\beta}=\frac{R^2 -\Lambda^2}{R^2+\Lambda^2}\,, \quad \sin{\beta}=\pm \frac{2R\Lambda}{R^2+\Lambda^2}\,,\vspace{-3.mm}
\end{equation}
\begin{equation}\label{vec_nRZ}
 \sin{\alpha}=\frac{Y}{R}\,,\quad \mbox{and}\quad \cos{\alpha}=\frac{X}{R}\,;
 \end{equation}
and the topological density \ref{top_density} is\vspace{-1.mm}
\begin{equation}\label{rho_sigma}
\rho_{\mbox{\tiny T}}=\pm\frac{\Lambda^2}{\pi(R^2+\Lambda^2)^2}\,.\vspace{-2.mm}
\end{equation}

Substituting expressions \ref{vec_nRXY_} and \ref{rho_sigma} into formulas \ref{Zeeman}
and \ref{Hartree} leads to the fact that integral \ref{Hartree} converges and
is well defined for any finite
value of $\Lambda$, whereas integral \ref{Zeeman}, at any nonzero $\epsilon_{\rm Z}$,
diverges logarithmically for any finite $\Lambda$. The study$\,$\cite{by98}
shows that actually the skyrmion should be characterized
by two length scales: $\Lambda$ -- the scale controlling the skyrmion `core', and
some value $l_{\rm sk}\sim (e^2/l_B\kappa\epsilon_{\rm Z})^{1/2}$ as a scale
characterizing the decrease in density $\rho_{\mbox{\tiny T}}(R)$ on the `tail'
-- at $R\gg\Lambda$.
The divergent integral is cut off at length $l_{\rm sk}$ considered to be much larger
than $\Lambda$, so that the subsequent minimization
procedure by using $\Lambda$ as a variational parameter, gives
the Zeeman \ref{Zeeman} and Hartree \ref{Hartree} energies to logarithmic accuracy.\cite{by98}

\vspace{-2.mm}

\section{Equivalences for the spatial
derivatives of the spin-rotation matrix components}

The spinor rotation matrix is $\,$\cite{ll91}
$$
{\hat U}(\mbox{\boldmath $R$})=\begin{pmatrix}\vspace{1.mm}
\cos\!\frac{\beta}{2}\,e^{i(\alpha+\gamma)\!/2} & \sin\!\frac{\beta}{2}\,e^{-i(\alpha-\gamma)\!/2}\,
\vspace{1.5mm}\\
-\sin\!\frac{\beta}{2}\,e^{i(\alpha-\gamma)\!/2} & \cos\!\frac{\beta}{2}\,e^{-i(\alpha+\gamma)\!/2}\,
\end{pmatrix}.
$$
The choice of functions $\alpha(\mbox{\boldmath $R$})$, $\beta(\mbox{\boldmath $R$})$
and $\gamma(\mbox{\boldmath $R$})$ is determined by our goal to find the lowest energy
spin excitation. In particular, the dependence of angle $\gamma$ on coordinate
$\mbox{\boldmath $R$}$
cannot be ignored (i.e., for example, if considering it to be constant), even despite formal
non-participation of $\gamma$ in determining the direction of the 3D unit vector [see. Eqs.
\ref{unit_n} and \ref{vec_n}].

Indeed, the additional `$\vec{\,\Omega}^{(l)}\!\!{\hat \sigma}_l$' terms in Eq.
\ref{H_1omega}, appearing
due to the noncommutativity of the $\mbox{\boldmath $\nabla$}$ and
${\hat U}\!(\mbox{\boldmath $R$})$ operators, are equal to
$
 \displaystyle{-i{\hat U}^\dag\mbox{\boldmath $\nabla$}{\hat U}\!\equiv\!\!\!
 \sum_{\, l=x,y,z}\!\!\!\mbox{\boldmath $\Gamma$}{}^{(l)}\!(\mbox{\boldmath $R$})\,
 \hat{\sigma}_l}$, where\vspace{-.5mm}
$\qquad\begin{array}{r}
\;{}\;{}\qquad\displaystyle{\mbox{\boldmath $\Gamma$}{}^{(x)}\!=
\left(-\sin{\alpha}\,
  \mbox{\small\boldmath $\nabla$}\beta+{\sin{\beta}\cos{\alpha}}
  \mbox{\small\boldmath $\nabla$}\gamma\right)\!/2},\vspace{1.mm}\\
{}\;{}\,\,\;\qquad\displaystyle{\!\mbox{\boldmath $\Gamma$}{}^{(y)}=
\left(\cos{\alpha}\,
  \mbox{\small\boldmath $\nabla$}\beta+\sin{\beta}\sin{\alpha}
\mbox{\small\boldmath $\nabla$}\gamma\right)\!/2},\;{}\;{}\,{}
\end{array}$\vspace{-.5mm}

\noindent and\vspace{-.0mm}

$\begin{array}{r}\qquad\displaystyle{\!\mbox{\boldmath $\Gamma$}{}^{(z)}\!=(
\vphantom{\mbox{\boldmath $\Gamma$}{}^{(z)}}
\mbox{\small\boldmath $\nabla$}\alpha+
\cos{\beta}\cdot\mbox{\small\boldmath $\nabla$}\gamma)\!/2\,}.{}\,\;\;\quad\qquad\quad{}
\vspace{-0.mm}
\end{array}$

\noindent If we suppose $\gamma\!=\,$const, then $\Rot\!\mbox{\boldmath $\Gamma$}{}^{(z)}\!\equiv 0$
wherever $\alpha(\mbox{\boldmath $R$})$ is regular, which is considered to occur
at any $\mbox{\boldmath $R$}\neq 0$). This leads
only to a trivial case with zero topological density \ref{top_density}, i.e.to the ground state.
At the same time, if assuming $\gamma=\alpha(\mbox{\boldmath $R$})$, we
find out that: first,
the non-physical singularity of $\mbox{\boldmath $\Gamma$}{}^{(z)}(\mbox{\boldmath $0$})$
(emerging
due to uncertainty of the $\alpha$ value at the point $\mbox{\boldmath $R$}=0$,
where $\cos{\beta(\mbox{\boldmath $0$})}=-1$) is canceled; second, the combination $\sum_{l}\!
\left[\mbox{\boldmath $\Gamma$}{}^{(l)}\right]^2$ represents exactly the energy density
defined in the framework of the $O(3)$ NL$\sigma$ model \ref{sigma} (the latter is
presumably suitable for a macroscopic description of extensive large-scale spin
excitations); third, the functions $\alpha(\mbox{\boldmath $R$})$ and
$\beta(\mbox{\boldmath $R$})$ may be chosen regular at any finite
$\mbox{\boldmath $R$}$, see below Eqs. \ref{Omegan}.
So replacing
${\vec \Omega}$${}_{\tiny{\mbox{\boldmath $R$}}}^{(l)}
=\left.\mbox{\boldmath $\Gamma$}{}^{(l)}\right|_{\gamma=\alpha}$,
we obtain\vspace{-1.mm}
\begin{equation}\label{A_Omega}
{}\!{}\!{}\!\begin{array}{l}
\displaystyle{\Omega_{\mbox{\tiny{\boldmath $R$}},\mu}^{(z)}}\!=
\!\frac{1}{2}\left(\vphantom{{\Omega}_\mu^{(z)}}1+
\cos{\!\mbox{$\beta$}}\right)\partial_{\mu}\mbox{$\alpha$}\,,
\quad\mbox{and}\vspace{.5mm}\\
\displaystyle{\Omega_{\mbox{\tiny{\boldmath $R$}},\mu}^{(x)}}\!=
\!-{\frac{1}{2}\sin{\mbox{$\alpha$}}{\vphantom{{\Omega}_\mu^{(z)}}}}
  \,\partial_\mu\mbox{$\beta$}+\!\displaystyle{\frac{\sin{\!\beta}}{2}
  {\vphantom{{\Omega}_\mu^{(z)}}}\!
  \cos{\mbox{$\alpha$}}\,
  \partial_\mu\mbox{$\alpha$}},\vspace{1.mm}\\
  \displaystyle{\Omega_{\mbox{\tiny{\boldmath $R$}},\mu}^{(y)}}\!=
\!{\frac{1}{2}\cos{\mbox{$\alpha$}}{\vphantom{{\Omega}_\mu^{(z)}}}}\,
  \partial_\mu\mbox{$\beta$}+\displaystyle{\frac{\sin{\!\beta}}{2}
  {\vphantom{{\Omega}_\mu^{(z)}}}\!
  \sin{\!\mbox{$\alpha$}}\,
  \partial_\mu\mbox{$\alpha$}},
  \vspace{-1.5mm}
\end{array}
\end{equation}
where $\mu=X$ or $Y$. The
following identities take place for these values and their combinations determined by
formulae \ref{Omega+/-}: \vspace{-2.mm}
\begin{equation}\label{rot_Omegaz}\begin{array}{l}
\Rot{\vec \Omega}^{(z)}\!\equiv\! 2\Omega_X^{(x)}\Omega_Y^{(y)}-2\Omega_X^{(y)}\Omega_Y^{(x)}\qquad\qquad\vspace{1.5mm}\\
\qquad\quad\,\,\equiv\displaystyle{\sin\!\beta\cdot
  \left(\partial_X\beta\cdot
  \partial_Y\alpha-\partial_Y\beta\cdot
  \partial_X\alpha\right)}
\end{array}\vspace{-2.mm}
\end{equation}
and\vspace{-2.mm}
\begin{equation}\label{Omegapm}
\Omega_\pm^-\Omega_\mp^+\equiv \frac{1}{2}\left[\sum_{\mu=X,Y\atop l=x,y}
\!\!\!\left(\!{\Omega}^{(l)}_\mu\!\right)\!\!{\vphantom{\left({\vec \Omega}^{(l)}
\pm\frac{1}{2}\!\Rot{\vec \Omega}^{(z)}
\right)}}^2\right]\pm\frac{1}{2}\Rot{\vec \Omega}^{(z)}\vspace{-2.mm}
\end{equation}
(the subscript ...${}_{\mbox{\scriptsize\boldmath $R$}}$ is omitted). Using Eq.
\ref{sigma} we also find that \vspace{-1.mm}
\begin{equation}\label{Omegan}
\sum_{\mu=X,Y\atop l=x,y}
\!\!\!\left(\!{\Omega}^{(l)}_\mu\!\right)\!\!{\vphantom{\left({\vec \Omega}^{(l)}
\pm\frac{1}{4}\!\Rot{\vec \Omega}^{(z)}
\right)}}^2\equiv\displaystyle{\frac{1}{4}\left[\left(\partial_X\!{\vec{\, n}}\right)^2\!+\!
  \left(\partial_Y\!{\vec{\, n}}\right)^2\right]},\vspace{-2.mm}
\end{equation}
where $\vec{\, n}$ is the 3D unit vector presented by Eqs. \ref{vec_n} or
 \ref{vec_i}.



\begin{thebibliography}{99}
\bibitem{foot2} This statement is true if function $F$
does not depend on the interaction of electrons belonging to different domains.
\bibitem{foot3} The calculation in Eq. \ref{Coul_ground} can be performed using, for instance, the excitonic representation for
interaction operator \ref{CoulR}.\cite{di20} After subsequent subtraction of the positive background energy, the result is\vspace{-2.mm}
$$
\!{}\!\begin{array}{l}
E^{(0,\nu)}_{\rm int}\!(\!\mbox{\boldmath $R$})=\!{-\pi\!
\sum_{\mbox{\scriptsize \boldmath $q$}}\!{\tilde V}(ql_B)\left[\delta_{\nu,1}\vphantom{+\,l_B^2q^2\!/2}
+\delta_{\nu,2}\left(\,2\right.\right.}\qquad\vspace{0.mm}\\
\quad\qquad\qquad\qquad\qquad\quad\left.\left. +\,l_B^2q^2\!/2\right)\right]e^{-q^2\l_B^2\!/2}.
\end{array}\vspace{-3.mm}
$$
${\tilde V}(p)$ is defined by equation \ref{tildeV} in the text. The summation is transformed
by the integration: ${\small \sum}_{\mbox{\scriptsize \boldmath $q$}}...\!\to
\!(N_\phi/2\pi)\!\int\!...\,l_B^2d^2\!\mbox{\boldmath $q$}\,$ [$N_\phi$ is given by Eq. \ref{Nphi}].  In case parameter
$r_s\!\sim\! {\mathscr E}_{\rm C}\!/\omega_c\,$ is considered to
be small, then the result $E^{(0,\nu)}_{\rm int}$ above is asymptotically exact to the first order in $r_s$.
\end{thebibliography}
\end{document}